\documentclass{llncs}

\usepackage{graphicx}

\begin{document}

\title{Developing Artificial Herders Using \emph{Jason}}

\author{Niklas Skamriis Boss \and Andreas Schmidt Jensen \and
J{\o}rgen Villadsen\thanks{Contact: \email{jv@imm.dtu.dk}}}

\institute{Department of Informatics and Mathematical Modelling \\
Technical University of Denmark \\
Richard Petersens Plads, Building 321, DK-2800 Kongens Lyngby, Denmark}

\maketitle \pagestyle{plain} \thispagestyle{plain}

\medskip

\begin{abstract}
This paper gives an overview of a proposed strategy for the ``Cows and
Herders'' scenario given in the Multi-Agent Programming Contest
2009. The strategy is to be implemented using the \emph{Jason} platform,
based on the agent-oriented programming language Agent\-Speak. The paper
describes the agents, their goals and the strategies they should
follow. The basis for the paper and for participating in the contest
is a new course given in spring 2009 and our main objective is to show
that we are able to implement complex multi-agent systems with the
knowledge gained in an introductory course on multi-agent systems.
\end{abstract}

\medskip

\section{Introduction}
This paper describes the work with a multi-agent system consisting of
artificial herders attempting to catch cows. The agents will compete
in the Multi-Agent Programming Contest 2009 (the scenario ``Cows and
Herders''). One of our main objectives in the contest has been to gain
experience with the development of multi-agent systems using \emph{Jason}.

Our basis for participating in the contest is the course ``Artificial
Intelligence and Multi-Agent Systems'' given in spring 2009 at the
Technical University of Denmark. The course provides an introduction
to multi-agent systems using \emph{Jason} as the implementation
platform. We hope to show that this introduction is sufficient to be
able to implement a more complex multi-agent system, such as the
``Cows and Herders'' scenario given in the contest.

\section{System Analysis and Design}
Our system consists of three kinds of agents: a herder, a scout and a leader.
The leader and the scout are basically herders with extra responsibilities.
The scout will initially explore the environment and subsequently act as an
ordinary herder. The leader will delegate targets to each of the herders --
including himself.

\newpage

Our system was designed using the Prometheus methodology as a guideline.
By this we mean that we have adapted relevant concepts from the methodology,
while not following it too strictly (as stated in \cite{Padgham+2007}). It has allowed
us to quickly identify the goals and what agents are needed to complete them.

\begin{figure}[htbp]
\centering
\includegraphics[height=2.5in,bb=0 0 580 322]{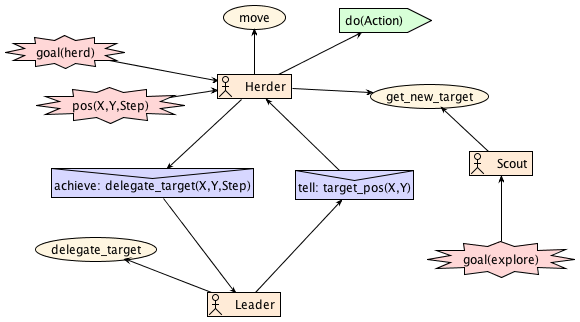}
\caption{Overview of the system.}
\label{fig:diagram}
\end{figure}

Figure~\ref{fig:diagram} gives an overview of the system. The diagram distinguishes
between the three types of agents, even though the leader and the scout are actually
special cases of the herder. This has been done to easily see the different roles each
agent plays. All agents know their own position and how many steps of the match
that have elapsed.
This is used to revise targets, since we do not want them to blindly follow a target.
An agent gets a new target by fulfilling the goal \texttt{get\_new\_target}.
The herders will tell the leader to delegate a target based on the agents current
position, while the scout will autonomously decide where to go.

We distinguish between the following types of targets. While the agents
do not really have an understanding of each concept, it is helpful for us to be
able to tell the targets apart.

\begin{description}
\item[Exploration targets] are targets in an area which has yet to be explored.
Such target is delegated to the scout, when he has not explored the entire
environment, or a herder, whenever he does not fulfill the criteria for a receiving
another type of target.
\item[Formation targets] are targets behind cows, but within a certain distance
from both cows and other herders, so that the group of cows can be controlled
and moved (or herded) towards the corral.
\newpage
\item[A switch target] is a target next to a switch. The reason for this is that an
agent should stand next to a switch in order to trigger it. This target will be
delegated whenever an agent is near a closed switch and it is reasonable to open it.
This is the case if one or more cows are near the fence or another
agent is on one side, while having a target on the other side (thus needing another
agent to open the fence, since one agent cannot pass a fence alone).
\end{description}

The scenario is quite dynamic since cows are continuously moving and fences
can be opened and closed, and all of this must be taken into account.

\section{Software Architecture}
Our strategy and agents are implemented using the \emph{Jason}
platform, which is an implementation of the Agent\-Speak language,
written in Java. \emph{Jason} is an effective platform for creating
multi-agent systems with a variable number of agents. Combined with
internal actions, we have a strong foundation for building a
multi-agent system, which not only uses the features of logic
programming, but allows us to develop imperative extensions as well.

The use of custom architectures in \emph{Jason} allows us to implement
a local simulation, as described in \cite{Bordini+2008}. This eases the
testing, as it can be done much faster.

As reference implementation we have used an implementation of the 2008 contest
made by the authors of \emph{Jason}. This has helped us getting started, even
though the scenario differs in many ways from last year.

Our solution to the contest was developed using Eclipse. The
implementation will have great focus on the advantages of object
oriented programming. This would also ease future expansion of more
agents etc. Shared memory could also be modelled by use of references
to shared objects used by multiple agents.

\section{Agent Team Strategy}
The agents will be moving around in a partially known environment. At the
beginning of a match everything is unknown, except for what lies within the
agents' field of view, and as the agents move around they gain knowledge
of the environment. The entire map is represented by a graph, where each
node in the graph represents a cell in the environment. When objects such
as obstacles or cows are discovered etc.\ the corresponding cell in the graph
will be assigned a value of that kind of object.

When agents move around they follow paths calculated by our navigation
algorithm. We have chosen to represent the environment as a graph, since it
makes it is easy to use a graph search algorithms for navigation. The actual
paths are calculated using the A* algorithm, which basically is an
advanced best-first search as it uses a heuristic to guide the search
for optimal paths.

A part of our strategy is to try to keep clustered cows together. This means that
the agents will have to move around a group of cows to avoid splitting them up.
This is ensured be assigning weight to the different cell in the graph. By
assigning higher weights to cells occupied by cows and cells adjacent to cows,
agents will navigate around a cluster instead of through it. Obstacles are handled
slightly different. The algorithm is implemented so that it does not consider cells
containing obstacles as valid cells for a path. This ensures that agents do not
try to move through obstacles.

To optimize the movement of our agents the paths are continuously calculated.
This is done since all agents can add new knowledge of obstacles etc.\ to the
graph as they perceive the environment. This ensures that if one agent discovers
that a corridor is blocked, then the other agents will try to move around it to get to
their target.

Experiments have shown that it is more efficient to herd cows in groups. To
ensure this the leading agent makes great use of a clustering algorithm. The
algorithm works be examining the surroundings of each cow; adjacent cows
are grouped together.

The strategy for herding the cows will be taken care of by the leading agent.
The team leader will coordinate the herding, ensuring that the cows are fleeing
the right way and that an agent will open the fence at an appropriate time.
Our strategy is mainly towards maximizing our own score. This means that our
agents will not try to capture cows already being herded by the opponent
deliberately, but it might happen if the leading agent estimates that they are
the cows closest to the corral.

An agent's beliefs consist of what they perceive and what others tell them to
believe. Optimally, we would like that every agent knows the same, i.e. they all
have the same beliefs. Unfortunately, since agents can only see a limited area
of the environment, this is not directly possible.

To ensure that every agent knows the same, any new belief an agent perceives
is sent to every other agent. All beliefs are shared immediately, since it does not
create much overhead and it is more efficient to share it than consider whether it
should be shared. When an agent discovers a static obstacle, every agent should
know this, so that their navigation can be adjusted to this new knowledge.

If an agent fails to achieve a given goal then we will use the \emph{Jason} failure
handling feature. This is done by implementing a deletion event \texttt{-!g}, which
will be executed if a given plan fails \cite{Bordini+2007}. After recovering from a
failed plan, we will attempt to reintroduce the goal (\texttt{+!g}) again.

\section{Discussion}
Our strategy is quite dynamic because of our use of path finding and
clustering algorithms, which allow the herders to fulfill their goals in
any given scenario. However, some of the choices we have made are
made on assumptions which may prove to be mistaken when the
competition is held.

We have decided to have a maximum cluster size (i.e. limit the number
of cows in a single cluster), because we believe that the agents may
have a hard time herding larger clusters. This may not be true, though,
since it could be more efficient to herd as many cows as possible as long
as they are clustered.

To compute an optimal search it is important to move agents in patterns so that the
largest possible area is explored. For example, agents should never
move side by side towards the same location, since this would not
exploit the full potential of the agents' field of view. Likewise it
could prove useful to move agents in patterns that ensures that no cow
can remain undetected in the explored area. However, we need to carefully
design our algorithms so that they do not take too long time to compute,
since the duration of a turn is limited.

At the time of writing this article our implementation is complete. However,
the contest has been postponed until after the deadline of the article, so
we are unable to discuss the results. We have managed to play a single
training match against another team, which we won. This match gave us
an opportunity to see how our team plays against others.

Generally we are quite satisfied with our system, which is able to fulfill the
goals of the scenario. Our strategy with a single leader delegating targets
lead to a less autonomous approach, but the \emph{Jason} framework
has allowed us to easily implement agents with certain goals and a way
to implement plans for handling these goals.

\section{Conclusion}
As discussed our primary strategy will be to maximize our own score
rather than prohibiting the opposing team from scoring points. This
has been done by optimizing the search for cows and guiding the cows
into the corral by using cooperating agents. Likewise all agents will
take the positions of the opponents into account when choosing a target.

Throughout the project we have considered problems such
as navigation, search for objects using multiple start points, clustering,
cooperation between agents and multi-agent planning. All planning was
implemented using Agent\-Speak, while external algorithms such as A*,
our clustering algorithm and target delegation were implemented in Java.

Despite our limited experience with Agent\-Speak and programming
intelligent multi-agent systems, we have managed to implement a fairly
reasonable system, with agents which fulfill the goals of the contest.
The ability of \emph{Jason} to implement custom architectures was a great
help during the work.

\end{document}